# Speculative Evolution Through 3D Cellular Automata: Synthetic Bones

Amir Hossein Khazaei, Texas A&M University, USA

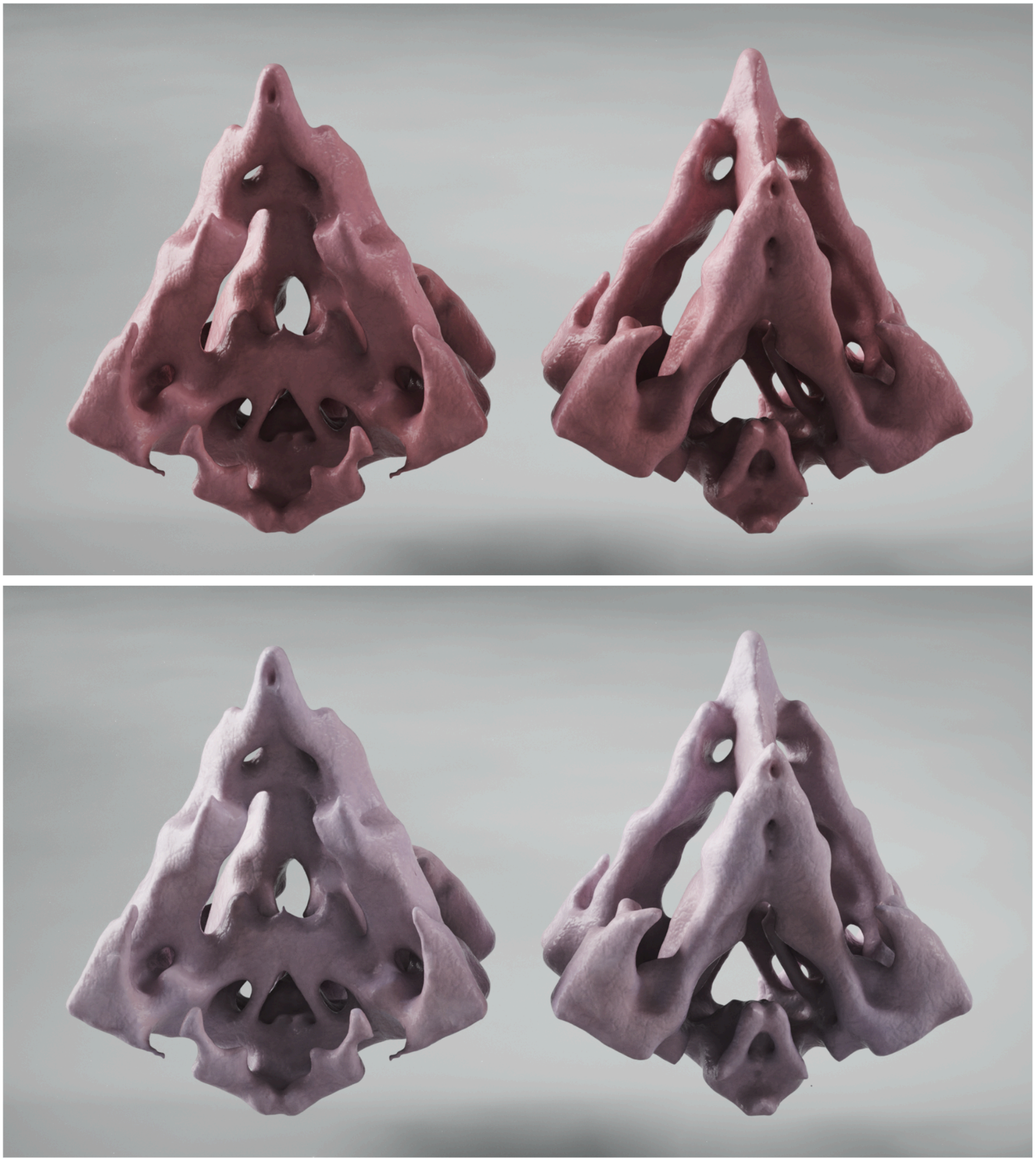

Fig. 1. Still Renders of the final Project/ Seed 49/ Iteration 60 that demonstrates the 3D implementation of Conway's Game of Life, using procedural simulation to generate unfamiliar (extraterrestrial) organic forms.

This project explores speculative evolution through a 3D implementation of Conway's Game of Life, using procedural simulation to generate unfamiliar (extraterrestrial) organic forms. By applying a volumetric optimized workflow , the raw cellular structures are smoothed into unified, bone-like geometries that evoke alien biological systems. The resulting forms, (strange yet organic) are 3D printed as fossil-like artifacts, offering a tangible look into hypothetical generated structures. This process blurs the line between artificial life, evolutionary theory, and digital fabrication, suggesting a new direction for computational bio-art. The work demonstrates how simple rules can simulate complex biological emergence and challenge perceptions of the organic.

*Both authors contributed equally to this research.

Author's Contact Information: Amir Hossein Khazaei, amirkhazaei@tamu.edu, Texas A&M University, College Station, Texas, USA.

## 1 Introduction & Related Works

The project draws inspiration from both computational science and speculative biology. Conway's Game of Life (1970) has long been a symbol of emergent complexity from simple rules, while its extension into three dimensions opens new opportunities for spatial and biological interpretation. Dougal Dixon's After Man (1981) laid the groundwork for speculative evolution as a genre, imagining post-human ecologies shaped by evolutionary pressures. In the digital arts, Refik Anadol's data sculptures and generative environments influence this project's aesthetic and conceptual foundation—particularly in translating data or simulations into spatial experiences. Ernst Haeckel's Art Forms in Nature (1904) serves as a historical precedent in biologically-inspired form-making, emphasizing symmetry, complexity, and the underlying logic of nature. Contemporary artists working in procedural sculpture and bio-art, such as Neri Oxman, further bridge organic systems with computational design. In this context, the project positions itself as a hybrid merging cellular automata with speculative zoology, and uniting digital life with tangible outcomes through 3D printing. It also echoes the educational goals of fictional biology projects like All Yesterdays (2013), where speculative forms not only entertain, but invite reflection on evolution, morphology, and the potential of unseen life forms, past or future.

## 2 Methodology

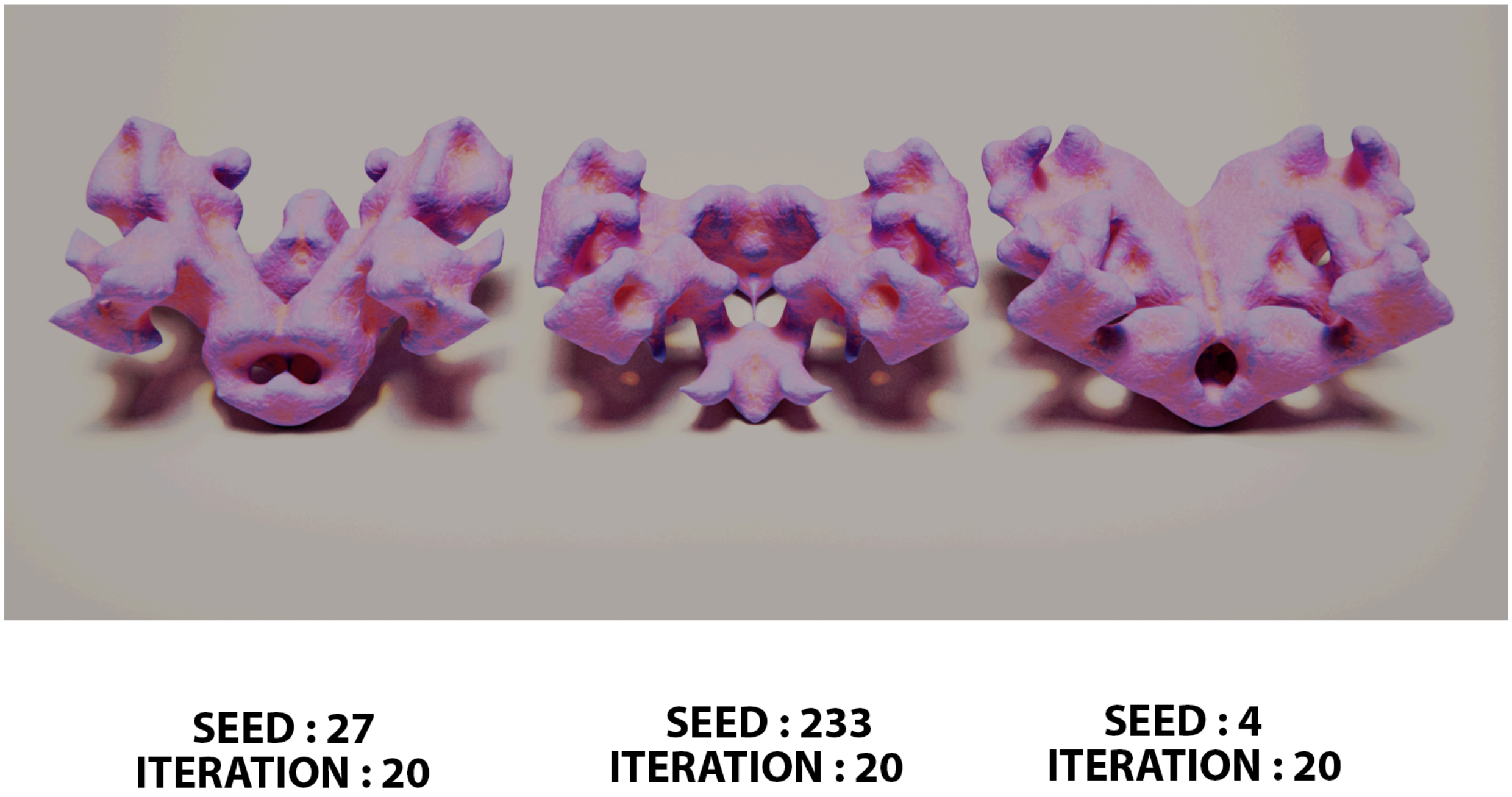

Fig. 2. The result of the generative pipeline with three different seeds and the same iteration(20)

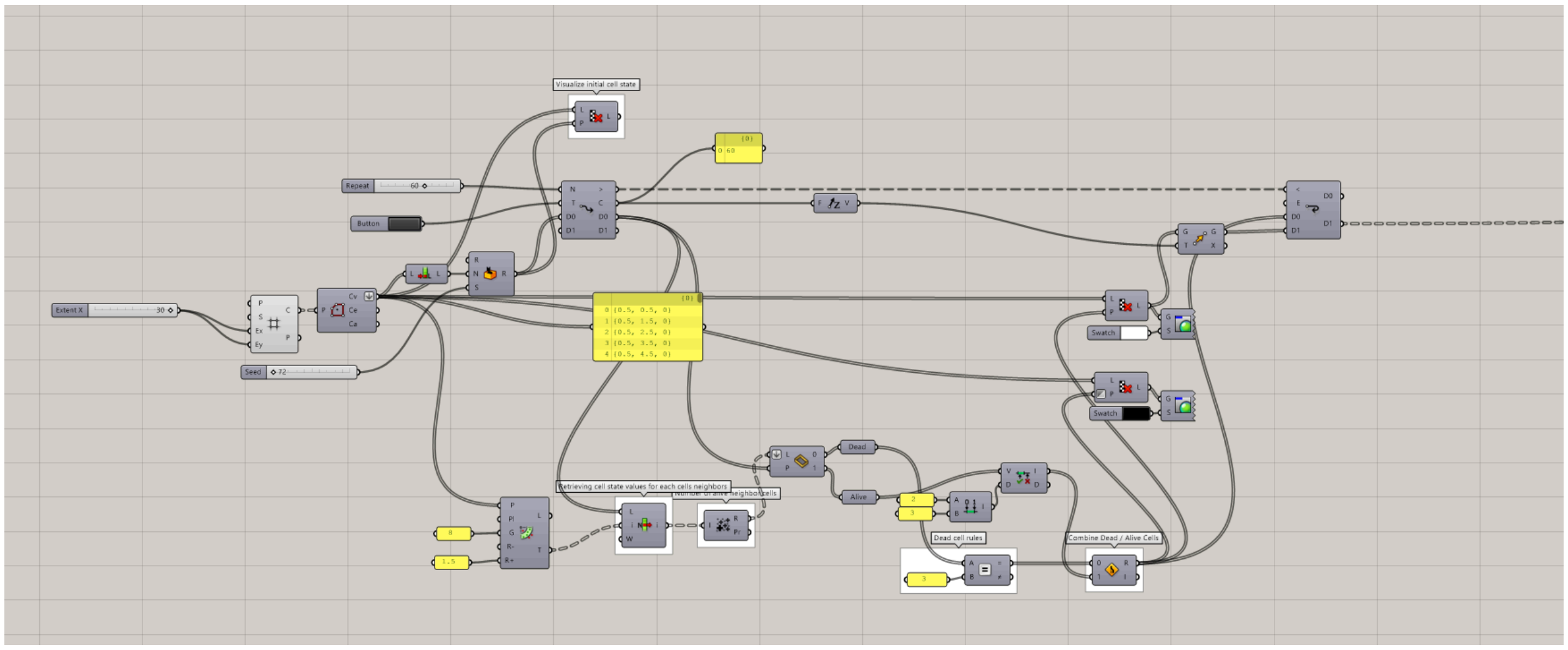

Fig. 3.VDB process operations such as smoothing, erosion, and dilation were used to transform the cubic automata into more cohesive and lifelike surfaces

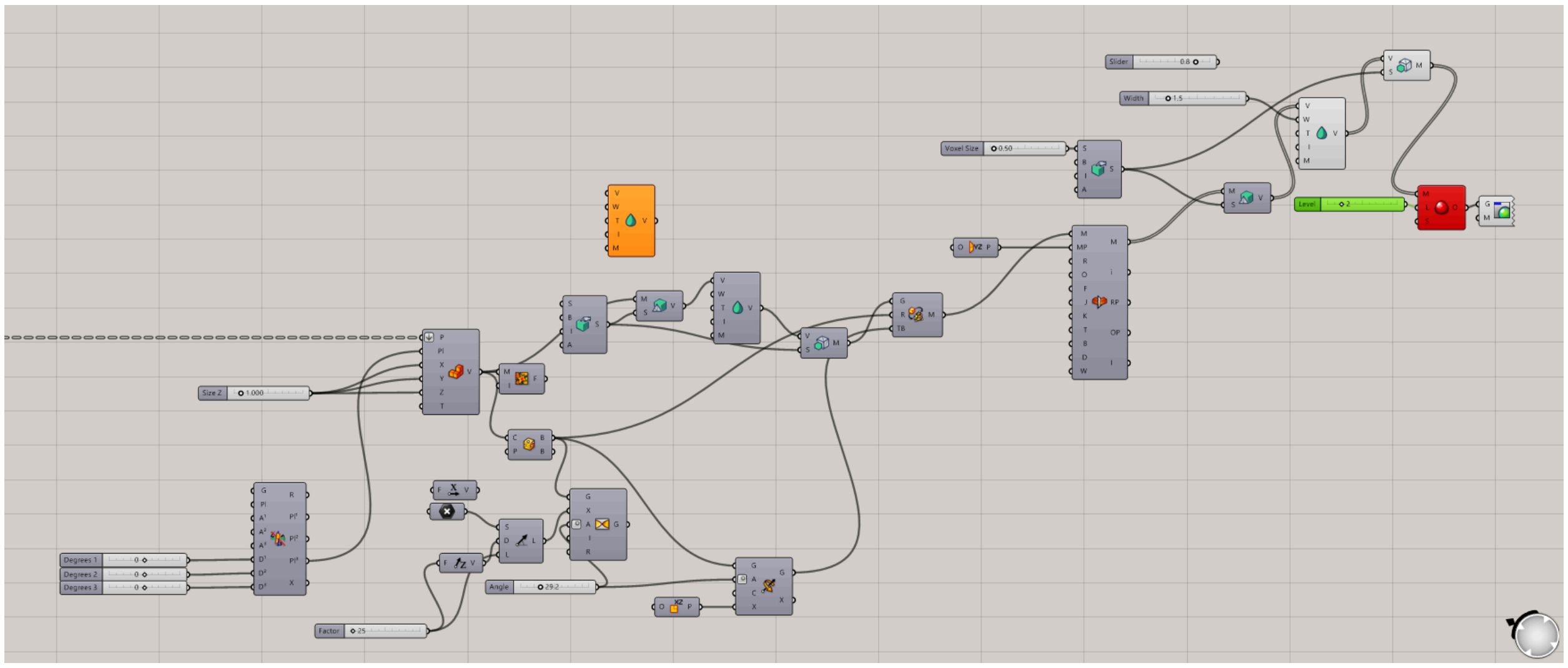

Figure 4. The initial 3D Cellular Automata Conway's game of life script that has been developed for the previous assignment

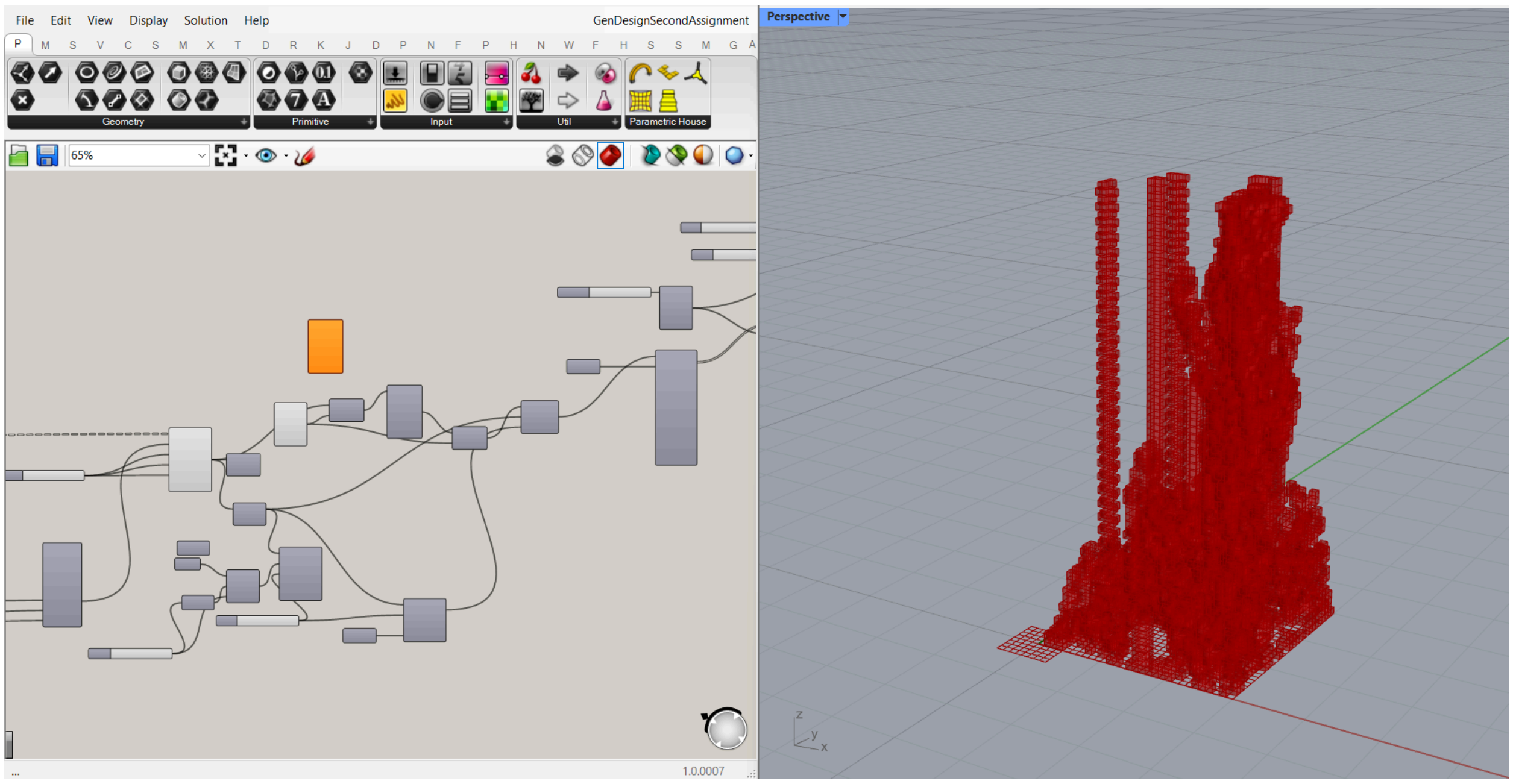

Figure 5. Grasshopper Script that demonstrates the result of the simulation with sed number 49 and 60 iteration (raw voxel cells)

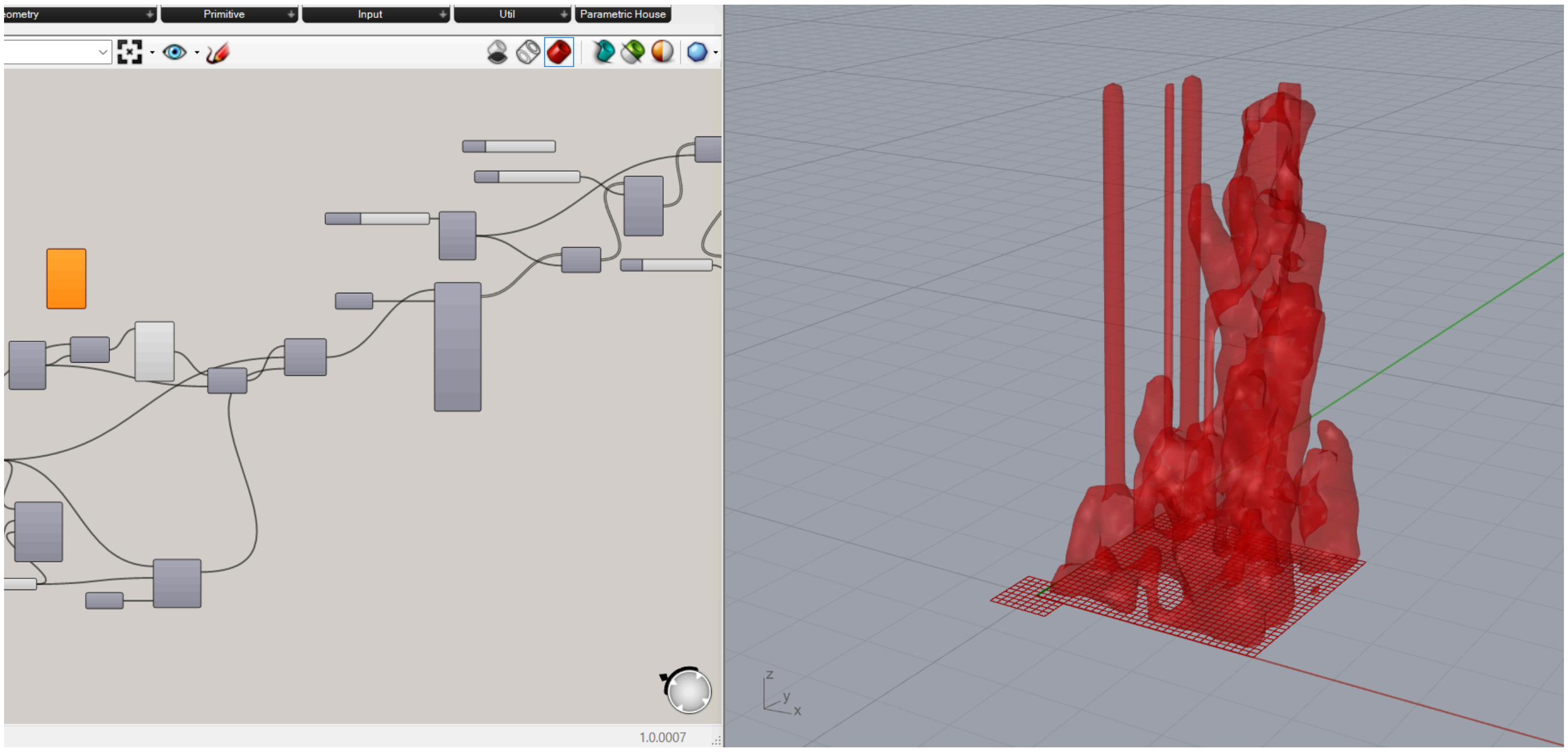

Figure 6. Grasshopper Script that demonstrates the smoothened result of the simulation with sed number 49 and 60 iteration (raw voxel cells)

The process begins with a custom 3D adaptation of Conway's Game of Life (an algorithmic system in which simple local rules lead to emergent global complexity). In contrast to its traditional 2D grid, the simulation operates within a voxelized 3D environment, allowing for the formation of spatial structures that more closely resemble organic growth. The rules were adjusted experimentally to encourage the persistence of complex forms, mimicking the behavior of biological structures.

Once the desired results with determined seeds and iterations were generated, the raw voxel data was optimized and by developing a script with different attributes like rotating the initial voxel cells, mirror cutting the final mesh (volumetric workflow using the VDB (Volumetric Data Block) system), applying smoothness and unifying the voxel cells. VDB operations such as smoothing, erosion, and dilation were used to transform the cubic automata into more cohesive and lifelike surfaces. This process enhanced the organic aesthetic while preserving the underlying logic of their formation.

The smoothed geometry were then prepared for 3D printing by converting form into high-resolution meshes and optimizing them for physical fabrication. The printed result exhibits forms that appear skeletal, coral-like, or alien fossils shape suggesting life forms shaped by unfamiliar evolutionary forces.

Throughout the process, the project maintained a speculative design lens, treating the generated artifacts not just as abstract sculptures, but as structures of fictional species from an alternate biology. This narrative context deepened the work's relationship to speculative evolution, turning algorithmic outputs into conceptual artifacts. The project was exploring the intersection of evolutionary biology, procedural design.

Mathematical Models used in the Grasshopper script: Anemone Loop: $X(n+1) = f(X(n))$ Proximity 2d Formula for finding the neighbours $PQ = d = \sqrt{[(x2 - x1)2 + (y2 - y1)2 + (z2 - z1)2]}$.

The smoothed geometry were then prepared for 3D printing by converting form into high-resolution meshes and optimizing them for physical fabrication. The printed result exhibit forms that appear skeletal, coral-like, or alien fossils shape suggesting life forms shaped by unfamiliar evolutionary forces.

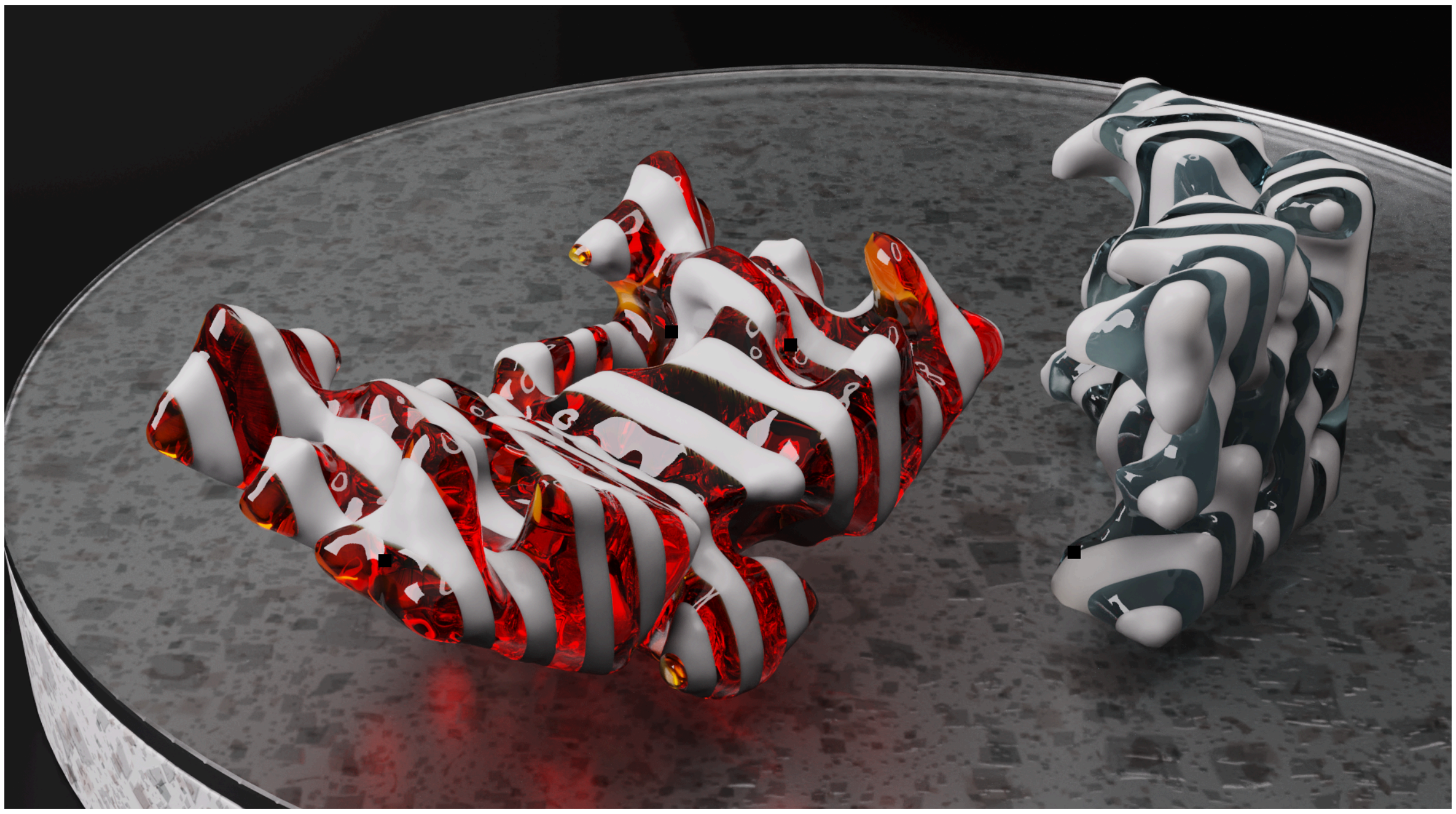

Figure 7. Still render that demonstrates the Contoured Geometry

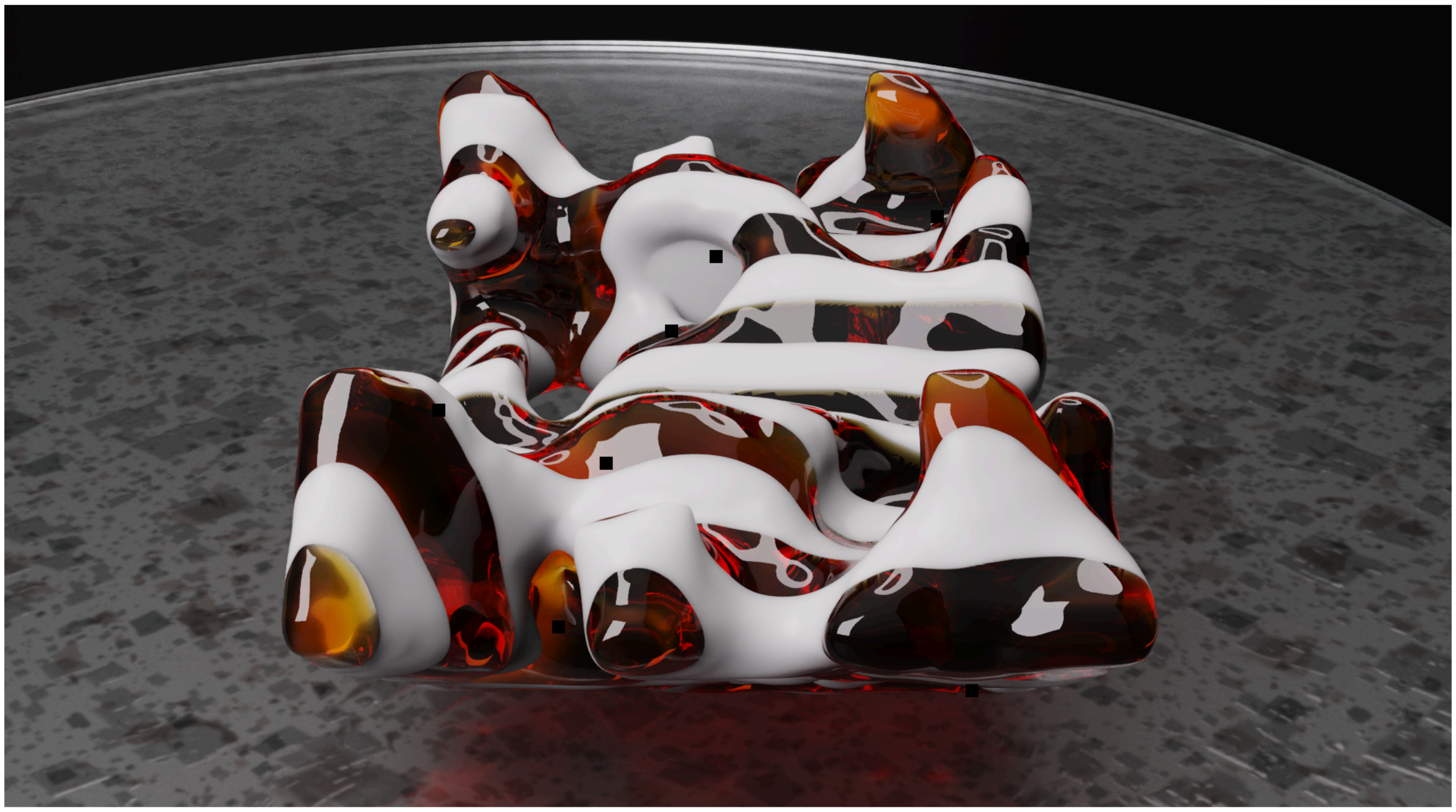

Figure 8. Still render that demonstrates the Contoured Geometry

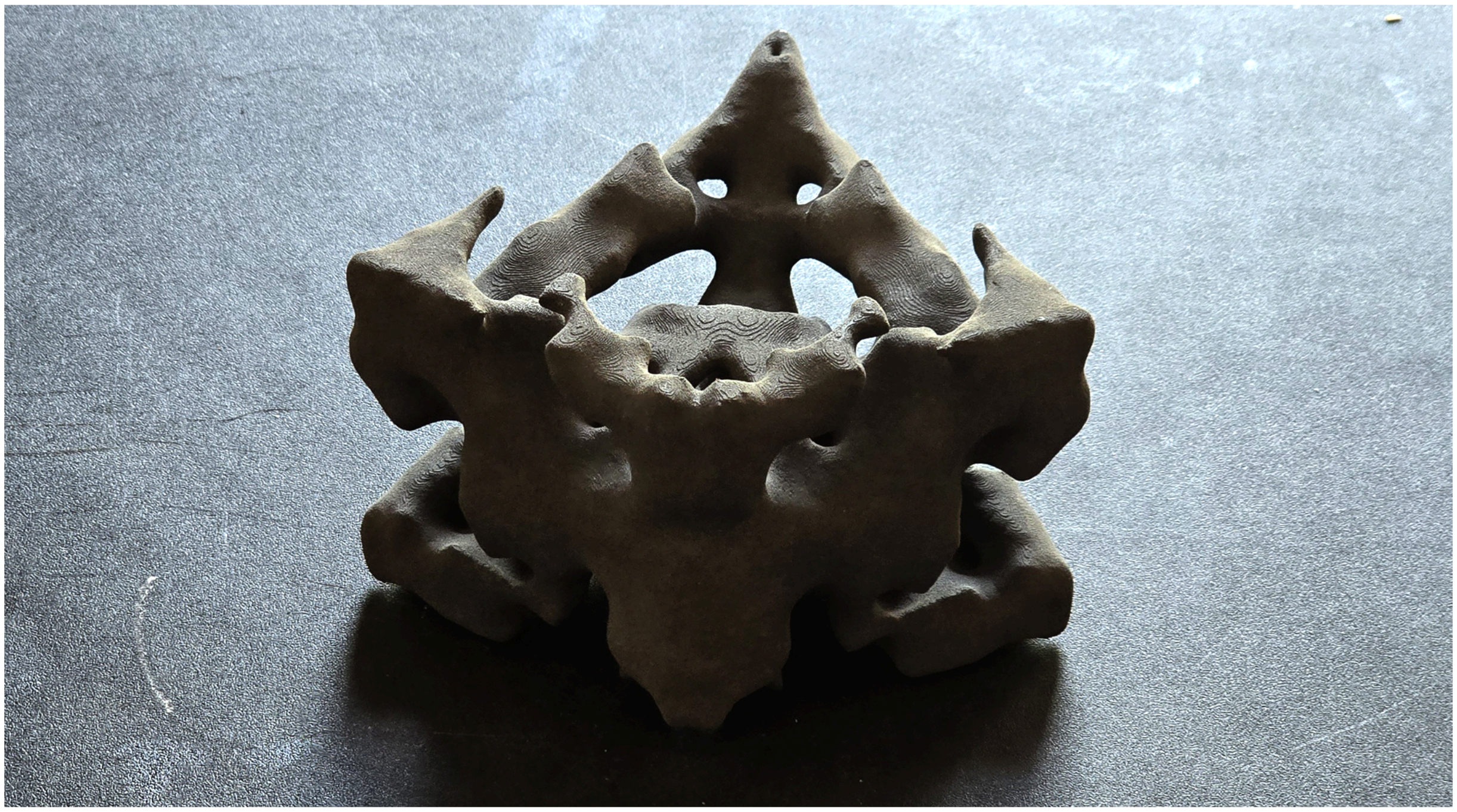

Figure 9. an image of the physical 3d printed form

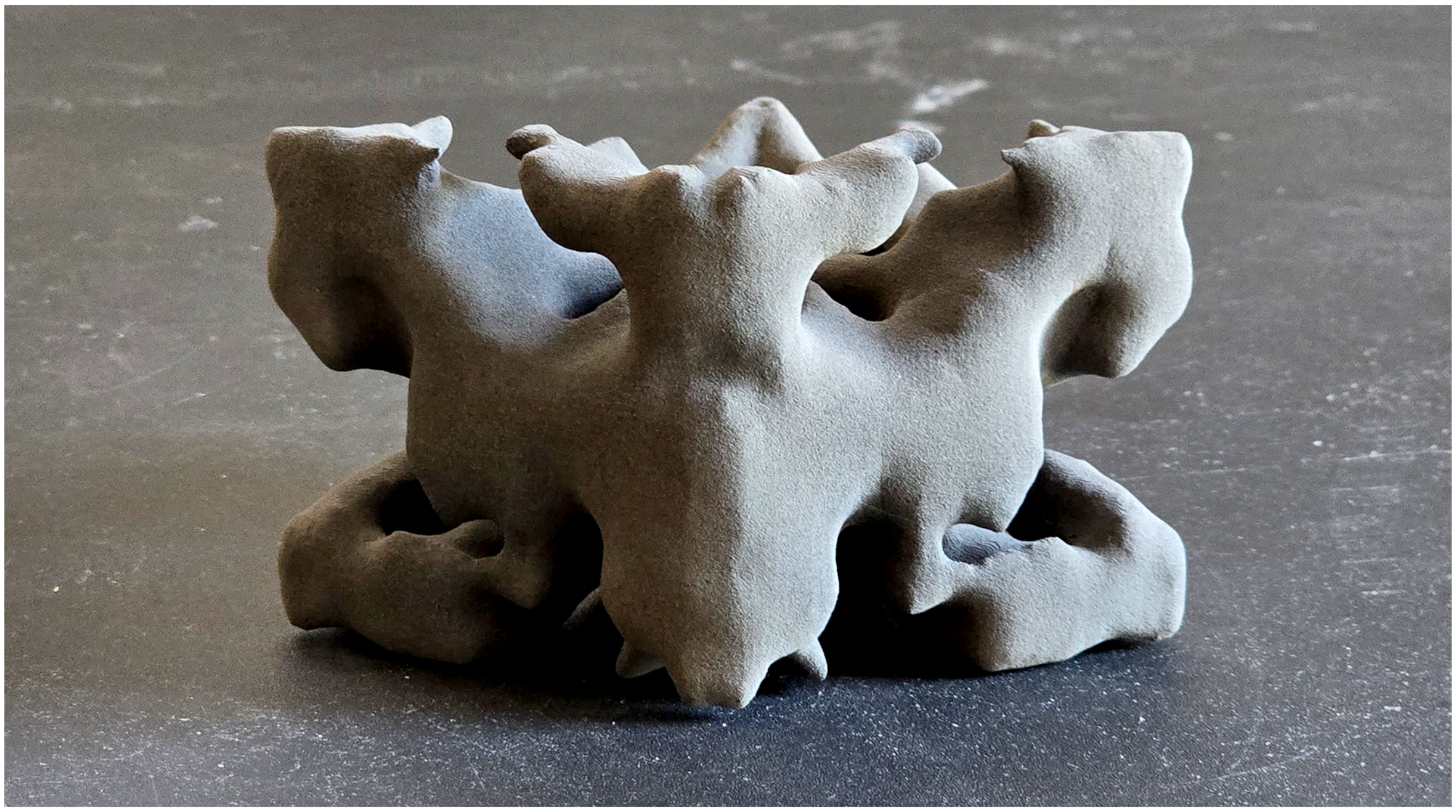

Figure 10. an image of the physical 3d printed form

## 3 Result and Future Work

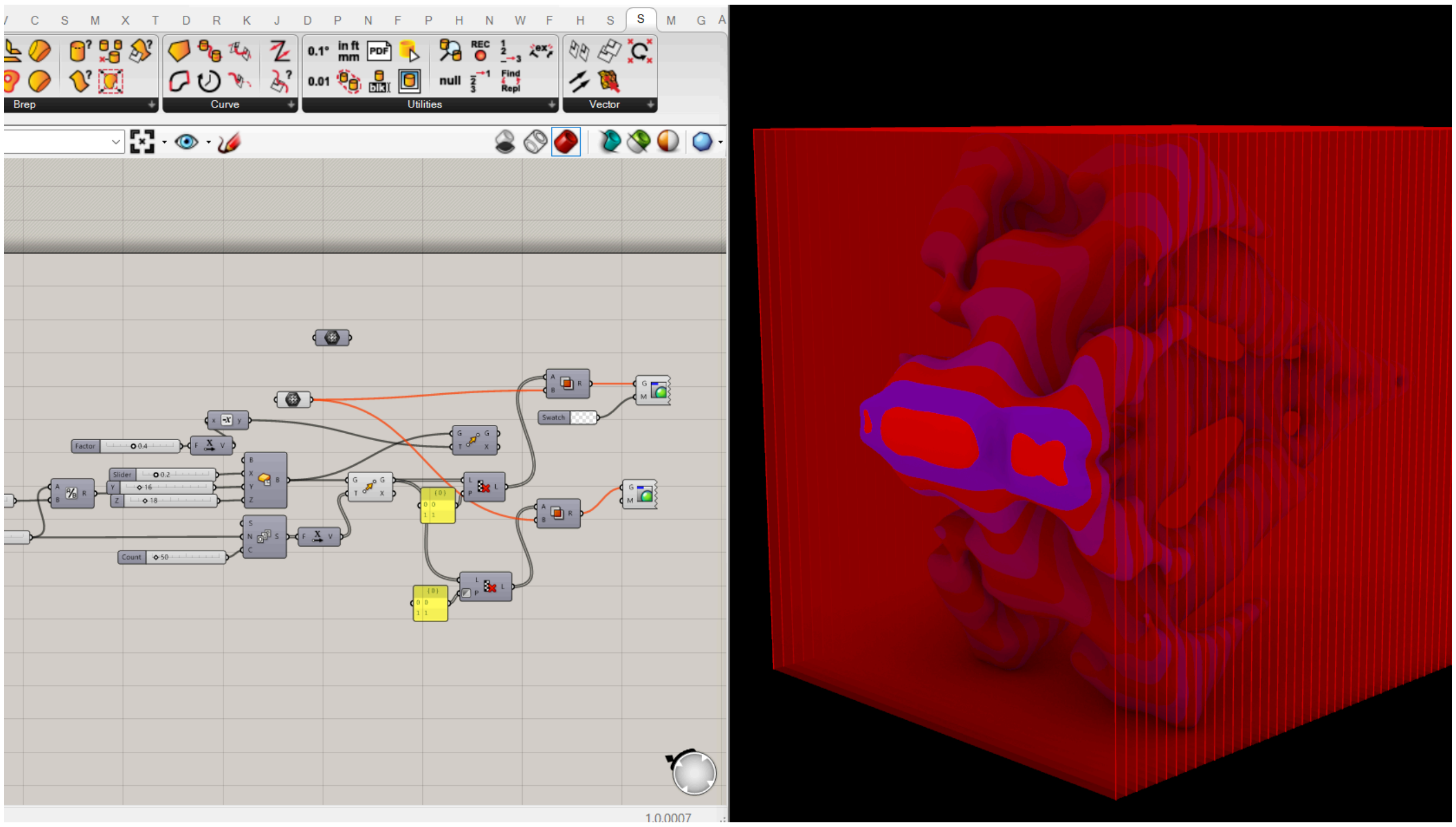

Figure 11. The script that demonstrates the process of mesh optimization and contouring the shape for 3d printing

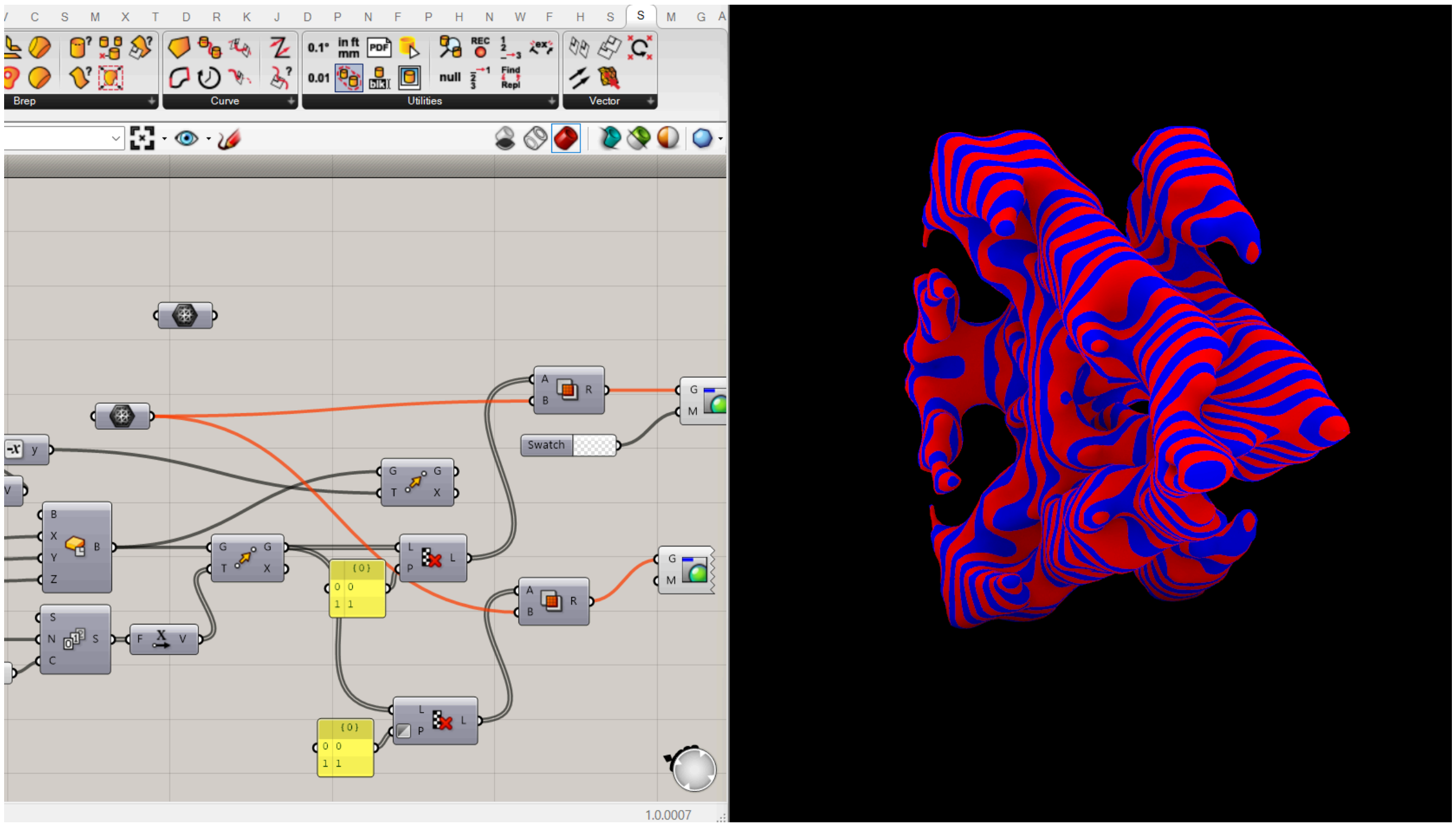

Figure 12. The script that demonstrates the process of mesh optimization and contouring the shape for 3d printing with the rhino view port

The final artifacts produced through this method are physically 3D printed sculptures that resemble abstract biological remains. Their unfamiliar forms provoke curiosity, bone spurs, or exoskeletal armor from speculative life forms. These shapes are rooted in simple algorithms yet suggest a deep evolutionary history. The project successfully demonstrates that algorithmic systems can serve as a generative base for speculative biology—creating forms that, while not real, feel plausible within a biological framework.

Future directions include expanding the simulation rules to mimic ecological interactions, such as predation or symbiosis, potentially leading to even more complex morphologies. A system could also be introduced to simulate mutation and natural selection, allowing the digital organisms to evolve over time based on survival heuristics. These developments would move the project toward a more dynamic and self-driven model of artificial life as well as exploring digital fabrication and exploring methods to print solid and transparent layers at the same time.

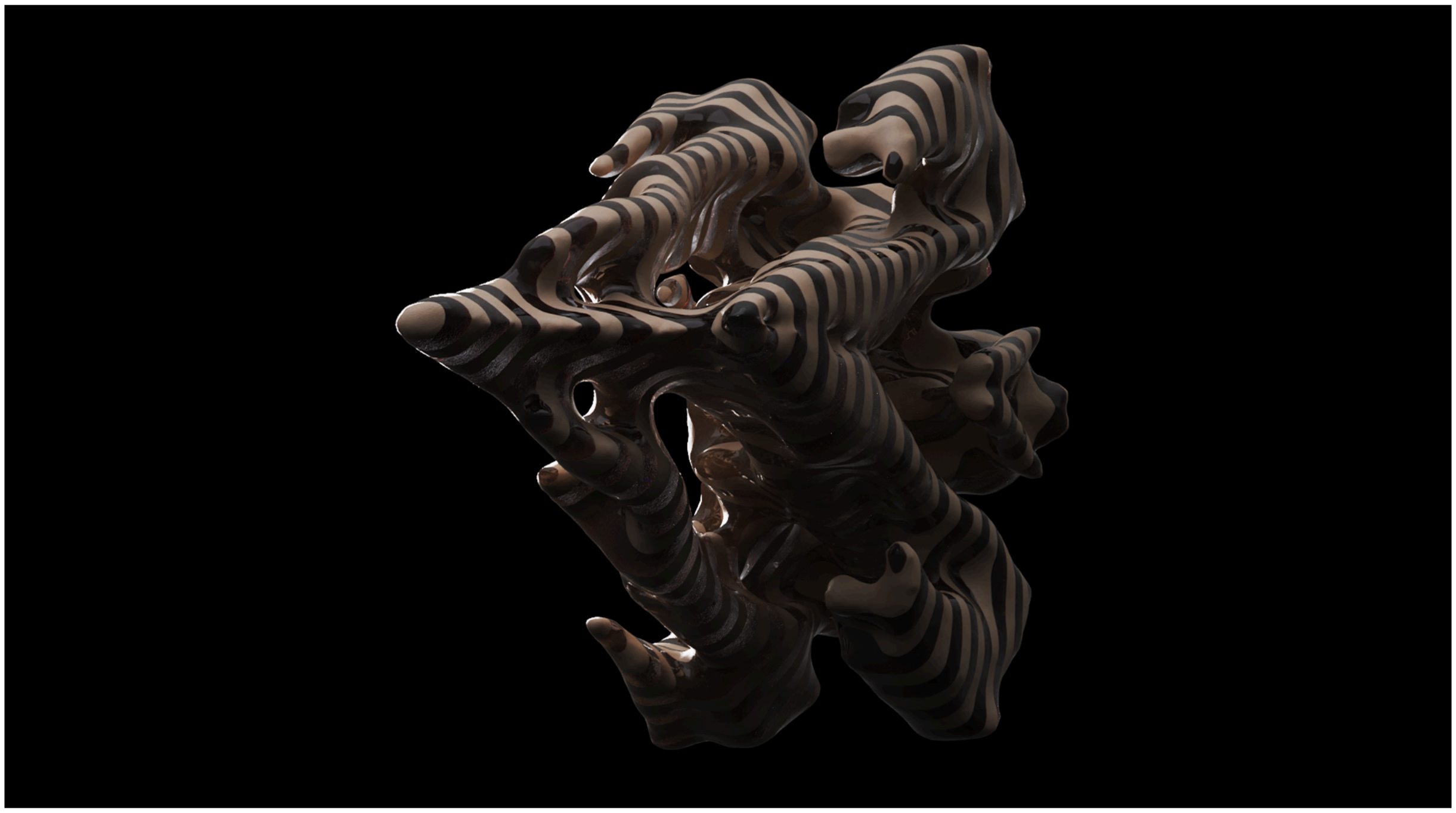

Figure 13. Still image demonstrates the contoured 3D Form for showing the layers with more details

## 4 Conclusion

This project demonstrates how simple computational rules can give rise to biologically suggestive and visually compelling forms. By combining 3D cellular automata, volumetric modeling, and digital fabrication, it transforms abstract data into tangible artifacts of fictional biology. Positioned within the speculative evolution genre, the work bridges art, science, and narrative—inviting reflection on how life could evolve under different conditions. It also showcases the potential of algorithmic design as both an artistic and scientific tool. These emergent forms not only provoke the imagination but also highlight the power of simulation to expand our understanding of life and complexity itself.

## Acknowledgments

This work is submitted as part of Assignment 1 for the VIZA 626 course at Texas A&M University,
under the instruction of Professor You-Jin Kim, during the Spring 2025 semester.